\documentclass[twocolumn,showpacs,aps,epsfig]{revtex4}

\ifx\pdfoutput\undefined
\usepackage{graphicx}
\else
\usepackage[pdftex]{graphicx}
\usepackage{epstopdf}
\fi
 \newcommand{\bq}{\begin{equation}}
 \newcommand{\eq}{\end{equation}}
 \newcommand{\bqn}{\begin{eqnarray}}
 \newcommand{\eqn}{\end{eqnarray}}
 \newcommand{\nb}{\nonumber}
 \newcommand{\lb}{\label}

\usepackage[center]{subfigure}
\usepackage{amsfonts}

\begin{document}

\title{ Homothetic Self-Similar Solutions of the Three-Dimensional Brans-Dicke Gravity}
 \author{G. A. Benesh\thanks{E-mail: greg$\_$benesh@baylor.edu} and 
  Anzhong Wang\thanks{E-mail: anzhong$\_$wang@baylor.edu} }
\affiliation{  Department of Physics, Baylor University, Waco, TX76798  } 

\date{\today}

\begin{abstract}

All  homothetic self-similar solutions of the Brans-Dicke scalar field in three-dimensional
spacetime with circular symmetry are  found in closed form.

 \end{abstract}


\maketitle

\noindent{\bf Introduction:} Recently, self-similar solutions of the Einstein 
field equations have attracted much attention, not only because they   can be 
studied analytically through simplification of the problem, but also because of
their relevance in  astrophysics  \cite{Car} and critical phenomena in  gravitational 
collapse \cite{Chop93,Wang01,Gun03}. In the latter case, as
in statistical mechanics \cite{Bar79,Go92}, the self-similar solutions represent 
the asymptotes of critical collapse in the intermediate regions.

In this Letter, we shall present all the  self-similar 
solutions of the Brans-Dicke scalar field in $2+1$ Gravity. It should be noted that
it is well-known that the Brans-Dicke  and Einstein's theories are related one to the other
by conformal transformations. However, these transformations do not preserve self-similarity,
as we show explicitly below. Thus, it is not clear how to get self-similar 
solutions of one theory from the other. 
Specifically,  we first define the homothetic self similarity  
(or self similarity of the first kind) in the Brans-Dicke theory, and
then write down the Brans-Dicke equations in terms of those self-similar variables. 
It is found that in some specific cases we can solve  the equations directly, but in 
general they are too complicated to be solved. In those cases, we use 
conformal transformations to write those equations in a form that we can solve 
in closed form.

\noindent{\bf  Field Equations:} 
The action in three-dimensional Brans-Dicke theory  takes the form \cite{BBL86},
 \bq
 \lb{2.1}
 S =  \int{\left(-\frac{1}{2\kappa}\phi R 
 + \frac{\omega}{\phi} \left(\nabla\phi\right)^{2} +
{\cal{L}}_{m}\right) \sqrt{-g} \; d^{3}x},
 \eq
where $R$ is the Ricci curvature scalar, $\omega$ the Brans-Dicke parameter, $\kappa$
the gravitational constant, and ${\cal{L}}_{m}$ the Lagrangain density of any matter field(s).
For the sake of convenience, in this Letter we  choose units such that
$\kappa = 1$. The variation of Eq.(\ref{2.1}) with respect to $g_{\mu\nu}$ and $\phi$ 
yields the field equations,
\bqn
\lb{2.2a}
& & R_{\mu\nu} - \frac{1}{2}g_{\mu\nu}R = \frac{1}{\phi}\left(T^{m}_{\mu\nu}
+ T^{\phi}_{\mu\nu}\right),\\
\lb{2.2b}
& & \Box^{2}\phi = \frac{1}{2(1 + \omega)} T^{m},
\eqn 
where $\Box^{2} \equiv g^{\alpha\beta}\nabla_{\alpha} \nabla_{\beta}, \; 
T^{m}_{\mu\nu}$ denotes the energy-momentum tensor (EMT),  
$T^{m} \equiv g^{\mu\nu}T^{m}_{\mu\nu}$, and $T^{\phi}_{\mu\nu}$ is given by
\bqn
\lb{2.3}
T^{\phi}_{\mu\nu} &=& \frac{2\omega}{\phi}\left(\phi_{,\mu}\phi_{,\nu}
- \frac{1}{2}g_{\mu\nu}\left(\nabla\phi\right)^{2}\right)\nb\\
& & + \left(\phi_{;\mu\nu} - g_{\mu\nu} \Box^{2}\phi\right),
\eqn
with $(\;)_{,\mu} \equiv \partial(\;)/\partial x^{\mu}$, and a semicolon, $(\;)_{;\mu}$,
denoting the covariant derivative.
In this Letter, we  consider only the case where $T^{m}_{\mu\nu} = 0$, in which 
Eqs.(\ref{2.2a}) and (\ref{2.2b}) can be written as
\bqn
\lb{2.4a}
R_{\mu\nu} &=& \frac{2\omega}{\phi^{2}}\phi_{,\mu} \phi_{,\nu}
+ \frac{1}{\phi} \phi_{;\mu\nu}, \\
\lb{2.4b}
\Box^{2}\phi &=& 0.
\eqn
The metric of three-dimensional spacetimes with circular symmetry can be 
cast in the form,
\bq
\lb{2.5}
ds^2 =  2 e^{2\sigma(u,v)} du dv - r^2(u,v) d\theta^2,
\eq
where  $(u,v)$ is a pair of null coordinates varying in the range
$(-\infty,\infty)$, and $\theta$  the usual angular coordinate with the
hypersurfaces $\theta = 0, \; 2\pi$ being identified. 
The metric is unchanged under the coordinate transformations,
\bq
 \lb{2.6}
 u = u(\bar{u}),\;\;\;\; v = v(\bar{v}).
\eq

Corresponding to the metric (\ref{2.5}) the field equations (\ref{2.4a}) take the form,
\bqn
\lb{2.10a}
 r_{,uu} - 2\sigma_{,u} r_{,u} &=& - \frac{r}{\phi^{2}}\left[\phi\left(\phi_{,uu}
- 2\sigma_{,u}\phi_{,u}\right) \right.\nb\\
& &  \left. + 2\omega \phi^{2}_{,u}\right],\\
\lb{2.10b}
 r_{,vv} - 2\sigma_{,v} r_{,v} &=& - \frac{r}{\phi^{2}}\left[\phi\left(\phi_{,vv}
- 2\sigma_{,v}\phi_{,v}\right) \right.\nb\\
& &   \left. + 2\omega \phi^{2}_{,v}\right],\\
\lb{2.10c}
r_{,uv} + 2r\sigma_{,uv} &=& - \frac{r}{\phi^{2}}\left(\phi\phi_{,vv}
+ 2\omega \phi_{,u}\phi_{,v}\right),\\
\lb{2.10d}
2r_{,uv} &=& - \frac{1}{\phi}\left(r_{,u}\phi_{,v} + r_{,v}\phi_{,u}\right),
\eqn
while the Klein-Gordon equation (\ref{2.4b}) is given by,
\bq
\lb{2.11}
 2\phi_{,uv} + \frac{1}{r}\left(r_{,u}\phi_{,v} + r_{,v}\phi_{,u}\right)
 = 0.
\eq

\noindent{\bf Homothetic Self-Similar Solutions:} 
Homothetic self-similar solutions are defined by the existence of a conform
Killing vector, $\xi^{\mu}$,
\bq
\lb{3.1}
{\cal{L}}_{\xi} g_{\mu\nu} = 2 g_{\mu\nu}.
\eq
From the above   it can be shown that
\bqn
\lb{3.5}
& & {\cal{L}}_{\xi} g^{\mu\nu} = -2 g^{\mu\nu}, 
\;\;\; {\cal{L}}_{\xi} R_{\mu\nu} = 0,\nb\\
& & {\cal{L}}_{\xi} R = -2 R, \;\;\; {\cal{L}}_{\xi} G_{\mu\nu} = 0.
\eqn
One can also show that for spacetimes described by the metric (\ref{2.5}), 
Eq.(\ref{3.1}) implies
\bq
\lb{3.2}
\sigma(u,v) = \sigma(z), \;\;\;\; r(u,v) = (-u)s(z),
\eq
where $z$ denotes the self-similar variable, defined by
\bq
\lb{3.3}
z \equiv \frac{v}{(-u)},
\eq
and the corresponding conformal Killing vector $\xi$ is given by
\bq
\lb{3.4}
\xi = u\partial_{u} + v\partial_{v}. 
\eq
Combining Eq.(\ref{3.5})  with the field equations (\ref{2.4a}), we find that
\bq
\lb{3.6}
{\cal{L}}_{\xi} \phi = \alpha \phi,
\eq
where $\alpha$ is an arbitrary constant. Eqs.(\ref{3.4}) and (\ref{3.6}) have the
solution,
\bq
\lb{3.7}
\phi(u,v) = (-u)^{\alpha}\varphi(z).
\eq
Substituting Eqs.(\ref{3.2}) and (\ref{3.7}) into Eqs.(\ref{2.10a})-(\ref{2.11}),
we obtain
\bqn
\lb{3.8a}
& & 
z^{2}s'' - 2z \sigma'(zs' - s) = - \frac{s}{\varphi^{2}}
       \left\{\varphi\left[z^{2}\varphi'' + 2(1-\alpha)z\varphi'
      \right.\right.\nb\\
& & \left. \;\;\;\;\;\;\;\;\;\;\;\;\;
      - \alpha(1-\alpha)\varphi - 2z\sigma'(z\varphi' 
      - \alpha\varphi)\right]\nb\\
& & \left.\;\;\;\;\;\;\;\;\;\;\;\;\;  
+ 2\omega\left(z\varphi' - \alpha \varphi\right)^{2}\right\},\\
\lb{3.8b}
& & 
s'' - 2 \sigma's' =  - \frac{s}{\varphi^{2}}
\left\{\varphi\left(\varphi'' -2\sigma'\varphi'\right)
 + 2\omega{\varphi'}^{2}\right\},\\
\lb{3.8c}
& & 
z s'' + 2s  \left(z\sigma'' +\sigma'\right) = - \frac{s}{\varphi^{2}}
       \left\{\varphi\left[z \varphi'' + \left(1-\alpha\right)\varphi'\right] \right.\nb\\
& & \left. \;\;\;\;\;\;\;\;\;\;\;\;\;
+ 2\omega\varphi'\left(z\varphi' - \alpha \varphi\right)\right\},\\
\lb{3.8d}
& & 
2z s''   = - \frac{1}{\varphi}
       \left\{\varphi'\left(zs' -s\right) + s'(z\varphi' - \alpha\varphi)\right\},\\
\lb{3.8e}
& & 
2z \varphi''  + \left(2z\frac{s'}{s} + \left(1-2\alpha\right)\right)\varphi'
- \alpha\frac{s'}{s}\varphi = 0,
\eqn
where a prime denotes ordinary differentiation with respect to $z$. 
From Eqs.(\ref{3.8a}) and (\ref{3.8b}) we find that
\bq
\lb{3.9}
2(1+\alpha)\sigma' + \left(1- \alpha - 2\alpha\omega\right)
\left(2\frac{\varphi'}{\varphi} - \frac{\alpha}{z}\right) = 0,
\eq
 which has the first integral,
\bq
\lb{3.10}
2(1+\alpha)\sigma + \left(1- \alpha - 2\alpha\omega\right)
\ln\left(\frac{\varphi^{2}}{z^{\alpha}}\right) = a_{0},
\eq
where $a_{0}$ is an integration constant. To solve the rest of equations
(\ref{3.8a})-(\ref{3.8e}), it is  convenient to consider the cases 
$\alpha = -1$ and $\alpha \not= -1$ separately.

\noindent{\bf Case  A) $\alpha = -1$:}
In this case, from Eq.(\ref{3.10}) we can see that we  need to 
distinguish two subcases: $\omega \not= -1$ and $\omega = -1$.

\noindent{\bf Case  A.1) $\alpha = -1,\;\;  \omega \not= -1$:}
When $\omega \not= -1$, Eq.(\ref{3.10})  yields  
\bq
\lb{3.11}
\varphi(z) = \frac{\varphi_{0}}{z^{1/2}}, \;\;\; \varphi_{0} \equiv 
\exp\left(\frac{a_{0}}{4(1+\omega)}\right).
\eq
Substituting the above into Eq.(\ref{3.8d}) we find
\bq
\lb{3.12}
4z^{2}s'' + s = 0,
\eq
which has a particular solution $s = s_{0}z^{1/2}$. To find a general solution 
of Eq.(\ref{3.12}) let us set $s(z) = s_{0}(z) z^{1/2}$, for which
Eq.(\ref{3.12}) becomes 
\bq
\lb{3.12a}
zs''_{0} + s'_{0} = 0.
\eq
The general solution is given by $s_{0}(z) = c_{0}\ln(z) + c_{1}$,
where $c_{0}$ and $c_{1}$ are two other integration constants. Thus,
the general solution of Eq.(\ref{3.12}) is given by,
\bq
\lb{3.13}
s(z) = z^{1/2}\left(c_{0}\ln(z) + c_{1}\right).
\eq
Inserting Eqs.(\ref{3.11}) and (\ref{3.13}) into Eq.(\ref{3.8b}), and then 
integrating the resulting equation, we obtain
\bq
\lb{3.14}
\sigma(z) = \frac{1+\omega}{8c_{0}}\left(c_{0}\ln(z) + 2c_{1}\right)\ln(z)
+ \sigma_{0},
\eq
where $\sigma_{0}$ is another integration constant. Thus, in the case $\alpha = -1$ and
$\omega \not= -1$ the general solution is given by
\bqn
\lb{3.14a}
s(z) &=& z^{1/2}\left(c_{0}\ln(z) + c_{1}\right),\nb\\
\sigma(z) &=& \frac{1+\omega}{8c_{0}}\left(c_{0}\ln(z) + 2c_{1}\right)\ln(z)
+ \sigma_{0},\nb\\
\varphi(z) &=& \frac{\varphi_{0}}{z^{1/2}}, \;\; (\alpha = -1 \not= \omega),
\eqn
where $c_{0}, \; c_{1}$ and $\sigma_{0}$ are integration constants.

\noindent{\bf Case  A.2) $\alpha = -1,\;\;  \omega = -1$:}
When $\alpha = -1$ and $\omega = -1$, Eq.(\ref{3.10}) is satisfied identically
for $a_{0} = 0$. As a result,   one can show that  Eqs.(\ref{3.8a}) and (\ref{3.8b}) 
reduce to the same equation,
\bq
\lb{3.15}
\frac{s''}{s} + \frac{\varphi''}{\varphi}
- 2\sigma'\left(\frac{s'}{s} + \frac{\varphi'}{\varphi}\right) - 2
\left(\frac{\varphi'}{\varphi}\right)^{2} = 0,
\eq
which, together with Eq.(\ref{3.8c}), gives the first integral,
\bq
\lb{3.16}
\sigma'(z) = \frac{c_{0}}{zs\varphi},
\eq
where $c_{0}$ is another integration constant. On the other hand, from Eqs.(\ref{3.8d})
and (\ref{3.8e}) we obtain
\bq
\lb{3.17}
s(z) = \frac{1}{\varphi}\left(c_{0}\ln(z) + c_{1}\right).
\eq
Inserting the above into Eq.(\ref{3.16}) we find 
\bq
\lb{3.18}
\sigma(z) = \ln\left(c_{0}\ln(z) + c_{1}\right) + \sigma_{0},
\eq
while from Eqs.(\ref{3.15})-(\ref{3.18}) we obtain
\bq
\lb{3.19}
\varphi(z) = \frac{c_{0}}{z^{1/2}\left(c_{0}\ln(z) + c_{1}\right)}.
\eq
Combining all the above together,  we obtain the following general solution,
\bqn
\lb{3.20}
\sigma(z) &=& \ln\left(c_{0}\ln(z) + c_{1}\right) + \sigma_{0},\nb\\
s(z) &=& \frac{z^{1/2}}{c_{2}}\left(c_{0}\ln(z) + c_{1}\right)^{2},\nb\\
\varphi(z) &=& \frac{c_{2}}{z^{1/2}\left(c_{0}\ln(z) + c_{1}\right)},\;\;
(\alpha = -1 = \omega),
\eqn
where $c_{0},\; c_{1} $  and $\sigma_{0}$  are all integration constants.

\noindent{\bf Case  B) $\alpha \not= -1$:} In this case, Eq.(\ref{3.10}) yields
\bq
\lb{3.21}
\sigma(z) =  \frac{1- \alpha - 2\alpha\omega}{2(1+\alpha)}
\ln\left(\frac{z^{\alpha}}{\varphi^{2}}\right) + \sigma_{0},
\eq
where $\sigma_{0} \equiv a_{0}/[2(1+\alpha)]$. To integrate the rest of 
Eqs.(\ref{3.8a})-(\ref{3.8e}), it is  easier to use the conformal 
transformation,
\bqn
\lb{3.22a}
\tilde{g}_{\mu\nu} &=& \phi^{2}g_{\mu\nu},\\
\lb{3.22b}
d\tilde{\phi} &=& \sqrt{2(\omega + 2)}\; \frac{d\phi}{\phi},
\eqn
which brings the action (\ref{2.1}) with ${\cal{L}}_{m} = 0$ into the form,
\bqn
\lb{3.23a}
S_{BD} &=&   \int{\left(-\frac{1}{2\kappa}\phi R + \frac{\omega}{\phi} 
\left(\nabla\phi\right)^{2}\right) \sqrt{-g} \; d^{3}x}\nb\\
\lb{3.23b}
&=& \int{\left(-\frac{1}{2\kappa}\tilde{R} +   \frac{1}{2}
\left(\nabla\tilde{\phi}\right)^{2}\right) \sqrt{-\tilde{g}}\;  d^{3}x} \nb\\
&\equiv& S_{E}. 
\eqn
Quantities with  tildes denote those calculated from the metric
$\tilde{g}_{\mu\nu}$, which is often referred to as the Einstein frame, while
the metric $g_{\mu\nu}$ denotes the Jordan frame. Although there is still
disagreement regarding  which frame is physical  \cite{FGN98}, we take the 
point of view that the two actions given by Eq.(\ref{3.23a}) represent two 
different theories, and the correctness of them 
should be determined by  observation. Therefore, until
such a determination is done, both of them are worth    investigating.

Writing the metric element in the Einstein frame as,
\bq
\lb{3.24}
d\tilde{s}^{2} = \tilde{g}_{\mu\nu}dx^{\mu}dx^{\nu}
= 2e^{\tilde{\sigma}(u,v)}dudv - \tilde{r}^{2}(u,v)d\theta^{2},
\eq
from Eqs.(\ref{3.2}), (\ref{3.7}), (\ref{3.22a}), and (\ref{3.22b}) we find
that
\bqn
\lb{3.25}
\tilde{\sigma}(u,v) &=&  \alpha\ln(-u) + \Sigma(z),\nb\\
\tilde{r}(u,v) &=& (-u)^{1+\alpha}S(z),\nb\\
\tilde{\phi}(u,v) &=& \Phi(z) + c \; ln(-u), 
\eqn
where
\bqn
\lb{3.26}
S(z) &\equiv& s(z)\varphi(z), \nb\\
\Sigma(z) &\equiv& \sigma(z) + \ln\left(\varphi(z)\right),\nb\\
\Phi(z) &\equiv& \sqrt{2(\omega + 2)}\ln\left(\varphi(z)\right), \nb\\
c &\equiv& \alpha \sqrt{2(\omega + 2)}.
\eqn
It is interesting to note that the solutions $\{\tilde{\sigma}, \tilde{r},
\tilde{\phi}\}$ given by Eqs.(\ref{3.25}) and (\ref{3.26}) are not self-similar, unless
$\alpha = 0$. In other words, conformal transformations do not preserve self-similarity.
However,   once   solutions $\left\{\Sigma(z), \; S(z),\; \Phi(z)\right\}$ are known
in the Einstein frame, we can use Eq.(\ref{3.26}) to find the solutions 
$\left\{\sigma(z), \; s(z),\; \varphi(z)\right\}$ in the Jordan frame via its inverse
\bqn
\lb{3.27}
\sigma(z) &=& \Sigma(z) - \frac{\Phi(z)}{\sqrt{2(\omega + 2)}}, \nb\\  
s(z) &=& S(z)e^{-\Phi(z)/\sqrt{2(\omega + 2)}}, \nb\\
\varphi(z) &=& e^{\Phi(z)/\sqrt{2(\omega + 2)}}.
\eqn 
It can be shown that in terms of $\Sigma, \; \Phi$ and $S$ the Einstein field 
equations,
\bq
\lb{3.28}
\tilde{R}_{\mu\nu}
= \tilde{\phi}_{,\mu} \tilde{\phi}_{,\nu},
\eq
and the Klein-Gordon equation $\tilde{\Box}^{2}\tilde{\phi} = 0$
in the Einstein frame can be cast in the form,
\bqn
\lb{3.29a}
& & z^{2}S'' -2\alpha z S' + \alpha(1+\alpha)S 
- 2\left(z\Sigma' - \alpha\right)\nb\\
& & \;\;\;\;\;\; \times \left[zS' - \left(1+\alpha\right)S\right]
= - S\left(z\Phi' - c\right)^{2}, \\
\lb{3.29b}
& & S'' -2\Sigma' S' = - S{\Phi'}^{2}, \\
\lb{3.29c}
& & z S'' -\alpha  S' + 2 S\left(z\Sigma'' +\Sigma'\right)   
= - S\Phi'\nb\\
& & \;\;\;\;\;\; \times \left(z\Phi' - c\right), \\
\lb{3.29d}
& & z^{2}S'' -\alpha  S'=0, \\
\lb{3.29e}
& & 2z\Phi'' +\left[\left(1-\alpha\right) + 2z\frac{S'}{S}\right]\Phi'
- c\frac{S'}{S} = 0.
\eqn
Eq.(\ref{3.29d}) has the general solution,
\bq
\lb{3.30}
S(z) = a z^{1+\alpha} + b,
\eq
where $a$ and $b$ are two integration constants.   
Combining Eqs.(\ref{3.29a}) and (\ref{3.29b}) we find that
\bq
\lb{3.31}
\Sigma'(z) = \frac{c}{1+\alpha}\Phi'
+ \frac{1}{2(1+\alpha)}\left(\alpha(1+\alpha) - c^{2}\right)\frac{1}{z},
\eq
which yields
\bqn
\lb{3.32}
\Sigma(z) &=& \frac{c}{1+\alpha}\Phi(z)
+ \frac{1}{2(1+\alpha)}\left(\alpha(1+\alpha) - c^{2}\right)\ln(z)\nb\\
& &  +
\Sigma_{0},
\eqn
with $\Sigma_{0}$ being a constant. Substituting Eq.(\ref{3.30}) into
Eq.(\ref{3.29e}), we obtain 
\bqn
\lb{3.33}
& & 2z\left(a z^{1+\alpha} + b\right)\Phi'' +\left[a(3+\alpha)z^{1+\alpha}
+ b(1-\alpha)\right]\Phi' \nb\\
& & \;\;\;\;\;\;\;\;\;\;\;\; 
- ac(1+\alpha) z^{\alpha} = 0.
\eqn
Depending on the values of $a$ and $b$, the above equation can have 
physically quite different solutions. Let us consider
  the three possible cases, (a) $a \not=0, \; b = 0$, (b)
$a =0, \; b \not= 0$, and (c) $ab \not=0$, separately.

\noindent{\bf Case  B.1) $ a\not= 0, \;\; b = 0$:}
In this case, Eq.(\ref{3.33}) reduces to
\bq
\lb{3.34}
2 z^{2}\Phi'' + (3+\alpha)z\Phi' - c(1+\alpha) = 0.
\eq
It can be shown that its associated homogeneous equation has the 
general solution,
\bq
\lb{3.35}
\Phi^{h}(z) = A z^{-(1+\alpha)/2} + B,
\eq
where $A$ and $B$ are two integration constants. One can also show that
Eq.(\ref{3.34}) allows the particular solution, $\Phi^{s}(z) = c 
\ln(z)$. Therefore, the general solution of Eq.(\ref{3.34}) is given by
\bq
\lb{3.36}
\Phi(z) = c \ln(z) + A z^{-(1+\alpha)/2} + B,
\eq
which, together with Eq.(\ref{3.32}), yields,
\bqn
\lb{3.37}
\Sigma(z) &=& \frac{1}{2(1+\alpha)}\left\{\left(c^{2} 
+ \alpha(1+\alpha)\right)\ln(z) \right.\nb\\
& & \left. + 2c\left(A z^{-(1+\alpha)/2} 
+ B\right)\right\} + \Sigma_{0}.
\eqn
Inserting Eq.(\ref{3.30}) with $b = 0$ and Eqs.(\ref{3.36}) and (\ref{3.37}) 
into Eqs.(\ref{3.29a})-(\ref{3.29c}),
we find that they are satisfied only when $A = 0$. Therefore, in the 
present case the general solution is given by
\bqn
\lb{3.38}
\Sigma(z) &=& \frac{1}{2(1+\alpha)}\left(c^{2} 
+ \alpha(1+\alpha)\right)\ln(z) +  {\Sigma}_{0},\nb\\
S(z) &=& a z^{1+\alpha}, \nb\\
\Phi(z) &=& c \ln(z) + B, \;\; (a \not=0, \; b = 0).
\eqn

\noindent{\bf Case  B.2) $ a= 0, \;\; b \not= 0$}: In this case, Eq.(\ref{3.33}) 
reduces to
\bq
\lb{3.34a}
2 z\Phi'' + (1-\alpha)\Phi'  = 0,
\eq
which has the general solution,
\bq
\lb{3.35a}
\Phi(z) = A z^{(1+\alpha)/2} + B.
\eq 
On the other hand, from Eq.(\ref{3.32}) we find,
\bqn
\lb{3.37a}
\Sigma(z) &=& \frac{c}{1+\alpha}\left(A z^{(1+\alpha)/2} + B\right)\nb\\
 & & + \frac{1}{2(1+\alpha)}\left(\alpha(1+\alpha)- c^{2}\right)\ln(z)  
 + \Sigma_{0}.
\eqn
However, as in Case B.1, the rest of the field equations
(\ref{3.29a}-(\ref{3.29e}) requires $A = 0$. Thus, the general solution 
in this case is given by
\bqn
\lb{3.38a}
\Sigma(z) &=& \frac{1}{2(1+\alpha)}\left(\alpha(1+\alpha)- c^{2}\right)\ln(z) 
+  {\Sigma}_{0},\nb\\
S(z) &=& b, \nb\\
\Phi(z) &=&   B, \;\; (a =0, \; b \not= 0).
\eqn

\noindent{\bf Case  B.3) $ a  b \not= 0$:}
In this case, the associated homogeneous equation of Eq.(\ref{3.33}) can be
cast in the form,
\bq
\lb{3.39}
\frac{\Phi''}{\Phi'} = - \frac{3+\alpha}{2z} 
+ \frac{b(1+\alpha)}{z\left(az^{1+\alpha} + b\right)},
\eq
which has the general solution,
\bq
\lb{3.40}
\Phi^{h}(z) = A \times \cases{  \tan^{-1}\left(\frac{z^{(1+\alpha)/2}}{\chi}\right) 
+ \frac{B}{A}, & $ ab > 0$,\cr
\ln\left|\frac{\chi + z^{(1+\alpha)/2}}{\chi - z^{(1+\alpha)/2}}\right| + \frac{B}{A},
& $ab < 0$,\cr}
\eq
where $\chi \equiv \sqrt{|b/a|}$, $A$ and $B$ are two integration constants. 
To find a particular solution of Eq.(\ref{3.33}), let us try the solution,
\bq
\lb{3.41}
\Phi^{s}(z) = c \ln(z) + \Phi_{0}(z).
\eq
Then, Eq.(\ref{3.33}) takes the form,
\bq
\lb{3.42}
{Y^{s}}' + \frac{a(3+\alpha)z^{1+\alpha} + 
b(1-\alpha)}{2z\left(az^{1+\alpha} + b\right)} Y^{s}
- \frac{bc(1+\alpha)}{2z^{2}\left(az^{1+\alpha} + b\right)} = 0,
\eq
where $Y^{s}(z) \equiv d{\Phi^{0}}(z)/dz$. Setting
\bq
\lb{3.43}
Y^{s}(z) = Y_{0}(z)\frac{z^{(\alpha - 1)/2}}{az^{1+\alpha} + b},
\eq
where $Y = {z^{(\alpha - 1)/2}}/{(az^{1+\alpha} + b)}$ is a particular solution 
of the associated homogeneous equation of Eq.(\ref{3.42}), we find that
Eq.(\ref{3.42}) reads
\bq 
\lb{3.44}
Y'_{0}(z) = \frac{1}{2}bc(1+\alpha)z^{-(\alpha + 3)/2},
\eq
which has the solution
\bq
\lb{3.45}
Y_{0}(z) = -bc z^{-(1+\alpha)/2}.
\eq
Combining Eqs.(\ref{3.43}) and (\ref{3.45}) we find that
\bq
\lb{3.46}
\Phi'_{0}(z) = - \frac{bc}{z\left(az^{1+\alpha}+b\right)},
\eq
which has a particular solution
\bq
\lb{3.47}
\Phi_{0}(z) = - c \ln(z) + \frac{c}{1+\alpha}\ln\left(az^{1+\alpha}+b\right).
\eq
Then, from Eqs.(\ref{3.40}), (\ref{3.41}), and (\ref{3.47}) we find that
the general solution of Eq.(\ref{3.33}) is given by
\bq
\lb{3.48}
\Phi(z) = \Phi^{h}(z) + \frac{c}{1+\alpha}\ln\left(az^{1+\alpha}+b\right).
\eq
Substituting this expression into Eq.(\ref{3.32}) we obtain $\sigma(z)$. Collecting 
all the results together, we find that the general solution in this case is 
given by
\bqn
\lb{3.49}
\Sigma(z) &=& \frac{c}{1+\alpha}\Phi^{h}(z)
+ \frac{c^{2}}{(1+\alpha)^{2}}\ln\left(az^{1+\alpha}+b\right)\nb\\
& & 
+ \frac{1}{2(1+\alpha)}\left(\alpha(1+\alpha) - c^{2}\right)\ln(z) +
\Sigma_{0},\nb\\
S(z) &=& a z^{1+\alpha} + b,\nb\\
\Phi(z) &=& \Phi^{h}(z) + \frac{c}{1+\alpha}\ln\left(az^{1+\alpha}+b\right),\nb\\
& &
\;\;\;\;\;\;\;\;\;\;\;\;\;\;\;\;\;\;\;\;\;\;\;\;\;\;\;\; (ab \not= 0),
\eqn
where $\Phi^{h}(z)$ is given by Eq.(\ref{3.40}).

Once we have all the self-similar solutions, it would be very interesting to study
their linear perturbations, following the line given in \cite{GG02,HW02} for massless
scalar fields in Einstein's theory of three-dimensional gravity, by paying
particular attention to  critical solutions and the possible dependence of the unstable
modes    on the Brans-Dicke parameter $\omega$. In the four-dimensional 
case, Choptuik obtained numerically a weak $\omega$-dependence of the black hole mass scaling
parameter $\gamma$   \cite{Chop93},
$$
M_{BH} \propto (p - p^{*})^{\gamma}.
$$ 
The scaling parameter $\gamma$ is related to the unstable modes ($\gamma = |k_{1}|^{-1}$).
In contrast,    
Chiba and Soda found $\gamma$ to be strongly  $\omega$-dependent \cite{CS96}.



\begin{thebibliography}{99}


\bibitem{Car}   B.J. Carr and A.A. Coley, Class. Quant. Grav. {\bf 16},  R31 (1999);
H. Maeda, T. Harada, H. Iguchi, and N. Okuyama, Prog. Theor. Phys. {\bf 108},  819
(2002); {\bf 110},  25 (2003);  B.J. Carr and C. Gundlach, Phys. Rev. {\bf D67},
024035 (2003).

\bibitem{Chop93} M.W. Choptuik, Phys. Rev. Lett. {\bf 70}, 9 (1993);
  ``{\em Critical Behavior in Massless
Scalar Field Collapse}," in {\it Approaches to Numerical
Relativity, Proceedings of the International Workshop on Numerical
Relativity}, Southampton, December, 1991, Edited by Ray d'Inverno
(Cambridge University Press, Cambridge, 1992);
  ``{\em Critical Behavior in Scalar Field Collapse},"
in {\it Deterministic Chaos in General Relativity}, Edited by D.
Hobill et al. (Plenum Press, New York, 1994), pp. 155-175.

\bibitem{Wang01} A. Wang, ``{\em Critical Phenomena in
Gravitational Collapse: The Studies So Far}," arXiv:gr-qc/0104073;
 Braz. J. Phys. {\bf 31},  188 (2001), and references therein.
 
\bibitem{Gun03}     C. Gundlach,  Phys. Rep. {\bf 376},  339  (2003); and
references therein.


\bibitem{Bar79} G.I. Barenblatt, {\em Similarity, Self-Similarity, and
Intermediate Asymptotics} (Consultants Bureau, New York, 1989),
pp.9-10.

\bibitem{Go92} N. Goldenfeld,   {\em Lectures on Phase Transitions and the
Renormalization Group} (Addison-Wesley Publishing Company, New
York, 1992), pp.301-302.





\bibitem{BBL86} J.D. Barrow, A.B. Burd, and D. Lancaster, Class. Quantum Grav.
{\bf 3}, 551 91986).

\bibitem{FGN98} V. Faraoni, E. Gunzig, and P. Nardone,  Fund. Cosmic Phys. {\bf 20}, 
 121 (1999) [arXiv:gr-qc/9811047].
 
\bibitem{GG02} D. Garfinkle and C. Gundlach, Phys. Rev. {\bf D66},  044015 (2002).


\bibitem{HW02} E.W. Hirschmann, A. Wang, and Y. Wu, Class. Quantum Grav.  
{\bf 21} 1791 (2004) [arXiv:gr-qc/0207121].


\bibitem{CS96} T. Chiba and J. Soda, Prog. Theor. Phys. {\bf 96},  567 (1996).
 
 \end{thebibliography}
\end{document}